\documentclass[conference]{IEEEtran}
\IEEEoverridecommandlockouts

\usepackage{cite}
\usepackage{amsmath,amssymb,amsfonts}
\usepackage{mathtools}
\usepackage{threeparttable} %
\usepackage{algorithmic}
\usepackage{graphicx}
\usepackage{xcolor}
\usepackage{url}
\usepackage{hyperref}
\usepackage{booktabs}
\usepackage{multirow}

\begin{document}
\title{Whisper Has an Internal Word Aligner}

\author{\IEEEauthorblockN{Sung-Lin Yeh$^*$\thanks{*Equal contribution.}, Yen Meng$^*$, Hao Tang}
\IEEEauthorblockA{\textit{Centre for Speech Technology Research, University of Edinburgh, UK}}}
\maketitle

\begin{abstract}
There is an increasing interest in obtaining accurate word-level timestamps from strong automatic speech recognizers, in particular Whisper.
Existing approaches either require additional training or are simply not competitive.
The evaluation in prior work is also relatively loose, typically using a tolerance of more than 200 ms.
In this work, we discover attention heads in Whisper that capture accurate word alignments and are distinctively different from those that do not.
Moreover, we find that using characters produces finer and more accurate alignments than using wordpieces.
Based on these findings, we propose an unsupervised approach to extracting word alignments by filtering attention heads while teacher forcing Whisper with characters.
Our approach not only does not require training but also produces word alignments that are more accurate than prior work under a stricter tolerance between 20 ms and 100 ms.\footnote{The source code is available at \href{https://github.com/30stomercury/whisper-char-alignment}{https://github.com/30stomercury/whisper-char-alignment}.}

\end{abstract}

\section{Introduction}

Word alignments between speech and text have broad applications, such as speech segmentation \cite{schulze2020joint,yu2023speech}, clinical applications \cite{yeung2015improving,zhou2024yolo,lian2024ssdm,romana2024automatic}, and automatic dubbing \cite{virkar2021improvements}.
However, not all automatic speech recognizers are able to produce accurate word alignments, sequence-to-sequence models (sometimes known as attention-based encoder-decoder models \cite{6af3452a28a04980b2b8f5eb48730d36}) being one of the examples \cite{chan2016listen,chiu2018state,dong2018speech}. 
Several attempts have been made to extract word-level timestamps from the attention maps of sequence-to-sequence models \cite{godard2018unsupervised,boito2019},
but segmenting words with attention remains challenging \cite{sanabria-etal-2021-difficulty}.
To achieve a low word error rate, it is in principle not necessary for an ASR model to maintain representations that accurately reflect alignments.
Not representing accurate alignments is a common problem that has plagued most end-to-end ASR systems \cite{zeyer2017ctc,shi2021timestamp,yao2023delay,huang2024less,schmitt2025conformer}.

Despite the difficulties, there is an ongoing effort in obtaining word-level timestamps from sequence-to-sequence models~\cite{lintoai2023whispertimestamped, wagner2024crisperwhisper}, especially for existing strong models such as Whisper \cite{radford2023robust}.
A common practice to produce alignments is to average all or a subset of attention maps and use dynamic time warping (DTW) to assign start time and end time to the tokens ~\cite{radford2023robust, wagner2024crisperwhisper, lintoai2023whispertimestamped}.
However, among hundreds of attention heads (e.g., 384 in Whisper \textit{medium}), it is unclear which ones, if any, represent alignments.
Moreover, wordpieces in the vocabulary of these strong models are often complete words.
Representations of larger size wordpieces tend to be more contextualized, resulting in fuzzier alignments, such as the one at the top of Figure \ref{fig:best-head}.
On the contrary, when a sequence of words is represented with wordpieces of smaller sizes, the output sequence is effectively longer, resulting in more entries in the attention maps.
We hypothesize that using smaller size wordpieces at the output, in particular characters, would encourage the ASR model to produce finer attention maps that are potentially more suitable for extracting word alignments.

In this work, we study the attention maps of Whisper and ask 1) if there exist attention heads that represent alignments between output tokens and input speech, despite not being trained to do so.
Furthermore,
we ask 2) if it is possible to obtain finer alignments from attention maps by replacing wordpieces of larger sizes with characters.
Note that the task is to obtain timestamps for the decoded words, not to obtain forced alignments with ground truth word sequences.

To answer those questions,
we first study attention maps in the decoder of Whisper that resemble word alignments using ground truth alignments.
We refer to these heads as the \textit{oracle heads}, as they depend on the ground truth words.
In addition, we allow the oracle heads to differ across samples and to come from any decoder layers.
We find that the word alignments produced by the oracle heads are encouragingly close to those from the Montreal Forced Aligner (MFA)~\cite{mfa}.
These attention maps produced by the oracle heads are also distinctively different from those produced by other heads.
Based on these findings, we propose a simple heuristic to filter attention heads that are likely to represent alignments.
To answer the second question, we study how attention maps change when different tokenizations of the same word sequence is provided to the decoder (similar to teacher forcing).
We find that the attention maps are indeed finer when replacing wordpieces of larger sizes with characters, even though Whisper is not trained with characters.

We evaluate the proposed approach on TIMIT, LibriSpeech, and AMI, and compare to Whisper's own approach and other recent approaches, such as CrisperWhisper~\cite{wagner2024crisperwhisper} and WhisperX~\cite{bain2023whisperx}.
We note that the 200 ms tolerance used in prior studies when evaluating word alignments (such as in \cite{bain2023whisperx,wagner2024crisperwhisper}) is relatively loose, barring these approaches from certain applications.
Instead, we provide a comprehensive comparison of prior approaches under the tolerance between 20 ms and 100 ms.
We show that our approach not only performs better in nearly all cases, but is also simple, not requiring any additional training. The approach can also be applied to other sequence-to-sequence ASR models.
Our finding adds to the ongoing debate about the relationship between attention and alignments~\cite{yang2022supervised, sanabria-etal-2021-difficulty}, 
showing the existence of accurate alignments in attention maps.

\section{Methods}
Sequence-to-sequence (seq2seq) models are not typically trained to produce word timestamps.
We adopt a three-step process that others~\cite{radford2023robust,wagner2024crisperwhisper}
have also taken to produce word alignments, where we 1) compute attention maps, 2) filter them, and 3) extract alignments from them with dynamic time warping (DTW). In particular, we propose an approach to identify attention maps that are closely related to word timestamps.

\subsection{Constructing attention maps}
\label{seq:methods:teacher-forcing}

We first describe how we construct attention maps from seq2seq models.
A seq2seq model consists of an encoder (denoted as $\text{Enc}$) and a decoder (denoted as \text{Dec}).
Given an input sequence $x_1, \dots, x_T$, autoregressive decoding amounts to running, for $k=1, \dots, K$,
\begin{align}
y_k, a_k = \text{Dec}(h_{1:T}, y_{0:k-1}),
\end{align}
where $h_{1:T} = \text{Enc}(x_{1:T})$, $y_k$ is the $k$-th predicted token, and $a_k$ is its corresponding cross-attention weights\footnote{For clarity, we only show a single attention head, but the discussion can be easily extended to multiple heads.}.
The sequence of tokens $y_1, \dots, y_K$ is later turned into a sequence of words $w_1, \dots, w_N$, for example, for ASR evaluation.

Since the tokenization of a word sequence is typically not unique \cite{sennrich2015neural,kudo2018subword}, we can take another tokenization $y'_1, \dots, y'_{K'}$ of the same word sequence $w_1, \dots, w_N$.
We then construct a different sequence of cross-attention weights
 with teacher forcing \cite{bengio2015scheduled}
using $y'_1, \dots, y'_{K'}$. The autoregressive decoding becomes, for $k=1, \dots, K'$,
\begin{align}
\_, a'_k = \text{Dec}(h_{1:T}, y'_{0:k-1}),
\end{align}
we disregard the output tokens with a dummy variable $\_$.

To produce word alignments, the decoder is first run with regular autoregressive decoding to get a predicted word sequence $\hat{w}_1, \dots, \hat{w}_N$.
The corresponding rows of attention weights are stacked to form the attention map, a matrix $A$.
Another option is to re-tokenize the predicted word sequence (e.g. in Table~\ref{tab:tokenization-example}) and to use attention maps produced by the new tokens, such as characters, with teacher forcing.
The attention map can be filtered and postprocessed to obtain word alignments.

\begin{table}
    \centering
    \caption{An example sentence tokenized by Whisper and by characters (where `\textvisiblespace ' is the space character).}
    \begin{tabular}{|p{0.25\linewidth} | p{0.6\linewidth}|}
    \hline
    transcript & She had your dark suit \\
    \hline
    wordpiece (default) &  [`She', `\textvisiblespace had', `\textvisiblespace your', `\textvisiblespace dark', `\textvisiblespace suit'] \\
    \hline
    character & [`S', `h', `e', `\textvisiblespace ', `h', `a', `d', `\textvisiblespace ', `y', `o', `u', `r', `\textvisiblespace ', `d', `a', `r', `k', `\textvisiblespace ', `s', `u', `i', `t'] \\
    \hline
    \end{tabular}
    \label{tab:tokenization-example}
\end{table}

\begin{figure}
    \centering
         \includegraphics[width=0.95\linewidth]{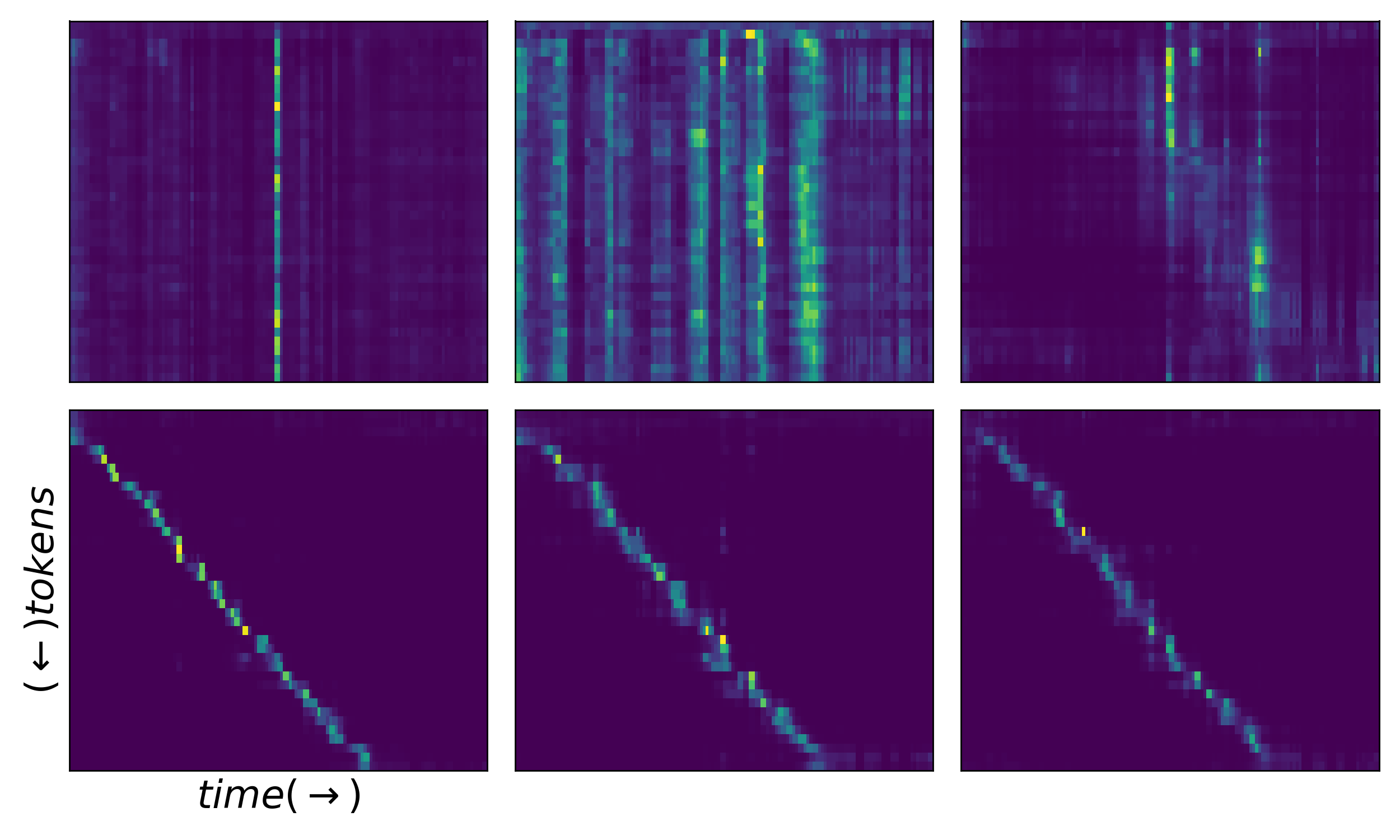} \\
    \caption{Example attention maps from Whisper.
    Alignments between the input and output are much more readily available from the attention maps in the second row than those in the first row.}
    \label{fig:different-heads}
\end{figure}

\subsection{Filtering attention heads}
\label{sec:method:head-select}

It is empirically common to see monotonic, alignment-like attention maps \cite{6af3452a28a04980b2b8f5eb48730d36}.
However, not all attention maps empirically look like alignments \cite{wang2024attention,schmitt2025conformer}.
In the case of Whisper, a few example attention maps are shown in Figure \ref{fig:different-heads}.
It is clear that some attention maps visually look like alignment while some clearly do not.

Several heuristics have been proposed to identify attention maps that look like alignments.
The entropy of attention weights is an intuitive measure \cite{wang2024attention}.
When an attention map looks alignment-like, each row of the attention map (i.e., the attention weights) needs to be concentrated on a particular word segment and thus has a relatively low entropy.
The coverage of input by the attention is another intuitive measure \cite{chorowski2016towards, klein2018opennmt}.
If an attention map represents an alignment, then every word segment should be attended and the coverage of the input should be high.
There are also attempts to measure diagonality of self-attention maps~\cite{zhang2021usefulness, yang2020understanding}, though they cannot be easily applied to cross attentions.

In this work, we propose to use the $\ell_2$ norm of columns and rows as a measure.
Formally, let the attention map $A$ be a matrix of $\mathbb{R}^{K \times T}$, where $K$ is the number of output tokens and $T$ is the number of input frames.
We define the score
\begin{align}
s(A) = \sum_{k=1}^K \|A_{k, \cdot}\|_2 + \sum_{t=1}^T \|A_{\cdot, t}\|_2,
\label{eq:norm_score}
\end{align}
where $A_{k, \cdot}$ is the $k$-th row of $A$, $A_{\cdot, t}$ is the $t$-th column of $A$, and $\|\cdot\|_2$ is the $\ell_2$ norm.
The $\ell_2$ norm serves a similar purpose as the Shannon entropy, measuring how concentrated the attention weights are.
To see this, recall that the R\'enyi entropy of order 2 of a discrete probability distribution $p$ is defined as $H_2(p) = -2 \log \|p\|_2$ and is a lower bound of the Shannon entropy.
Higher $\ell_2$ norm means that the distribution is more concentrated.
What is different in our approach is that we consider both $\ell_2$ norm of the rows and the columns.
The values need to be concentrated row-wise because ideally only a single word segment is attended.
The values need to be concentrated column-wise as well because ideally the same word should not be attended multiple times.

Once the top scoring attention maps are identified (with the hope that they look like alignments), we can then proceed to extract the actual word alignments.

\subsection{Extracting alignments}
\label{sec:method:dtw}

To obtain the timestamps of words, we follow~\cite{radford2023robust, wagner2024crisperwhisper}, constructing a cost matrix based on the attention maps and running DTW. DTW is a common algorithm for obtaining alignments; every token is mapped to one start time and one end time, allowing us to read off the timestamps at word boundaries.

As there are variations of DTW, we clarify our procedure below.
Given $H$ heads after filtering, the DTW computes the recursion,
for $i=1, \dots, N$ and $j=1, \dots, T$,
\begin{align}
 Q_{i,j} = \min (Q_{i-1, j}, Q_{i, j-1}, Q_{i-1, j-1}) - \bar{A}_{i,j} / \|\bar{A}_{:, j}\|_2,
    \label{eq:dtw}
\end{align}
where $A^h$ is the $h$-th attention map and $\bar{A} = \frac{1}{H}\sum_{h=1}^H A^h$.
Note that when computing the cost term $\bar{A}_{i,j} / \|\bar{A}_{:, j}\|_2$, the attention maps are first averaged over $H$ heads and then divided by the column norm.
The averaging gives some room for error when filtering the heads, and the performance should not hurt much as long as most of the heads look like alignments.
Given that the column norm behaves like entropy as we have discussed, the division of the norm has the effect of sharpening the attention weights.
Similar dynamic programming solutions have also been applied to extract monotonic alignments in \cite{kim2020glow,sanabria-etal-2021-difficulty,wagner2024crisperwhisper,yu2025unsupervised}.

\section{Experiments}

To study the two hypotheses 1) whether there exist attention heads that represent alignments and 2) whether it is possible to obtain finer alignments with characters, this section describes the datasets and baselines that we compare against.
Note again that the task is to obtain word timestamps for decoded word sequences, not to be confused with forced aligning ground truth word sequences to speech. 

\subsection{Datasets}

Our experiments are conducted on TIMIT \cite{garofolo1993darpa}, LibriSpeech \cite{panayotov2015librispeech}, and AMI \cite{renals2007recognition,carletta2005ami}.
TIMIT is the only data set that is phonetically transcribed.
Following \cite{wiegreffe-pinter-2019-attention}, we use the transcriptions from the TIMIT training set for evaluation. 
For LibriSpeech, we choose the dev-clean split, and use alignments from MFA \cite{mfa} (Kaldi GMM-HMM) as a proxy to the actual word alignments.
Following \cite{bain2023whisperx,wagner2024crisperwhisper}, we use the eval set from the individual headset microphones (IHM) of AMI.
How AMI is processed for this task is unfortunately unclear from \cite{bain2023whisperx,wagner2024crisperwhisper}.\footnote{Word timestamps generated by HTK are included in the AMI data set \cite{carletta2005ami}, but are unfortunately misread as manual annotations (e.g., in \cite{bain2023whisperx, wagner2024crisperwhisper}).}
We instead use a speaker-adapted GMM-HMM trained on the training set of AMI IHM (tri4a in Kaldi s5) to align the eval set.

\subsection{Baselines}

All Whisper experiments are conducted on the official checkpoints.\footnote{https://github.com/openai/whisper}
The official release produces word timestamps using the same general approach described in Section \ref{sec:method:dtw}, except that the heads are averaged over specific sets.
One option averages over the heads from the upper half of the decoder layers, while the other averages over a pre-defined, fixed set of heads.
We compare to CrisperWhisper \cite{wagner2024crisperwhisper}, a fine-tuned version of Whisper that follows the same approach to producing word alignments, except that only a fixed set of 15 heads (chosen based on TIMIT) are averaged.\footnote{https://github.com/nyrahealth/CrisperWhisper?tab=readme-ov-file\#5-how}
Note that we do not apply their pause heuristic to any of the approaches.
As in \cite{wagner2024crisperwhisper}, we strip punctuation in the transcript when aligning. 

Since TIMIT is phonetically transcribed, we present the results of MFA alignments evaluated on the phonetic transcriptions. This baseline is a sanity check of how close MFA alignments are to the transcriptions, allowing us to use MFA as a proxy for evaluating datasets without manual alignments.
For MFA, we use the acoustic model trained on 982 hours of LibriSpeech as in \cite{wiegreffe-pinter-2019-attention}.
We also include WhisperX \cite{bain2023whisperx}, which uses a fine-tuned wav2vec 2.0 for ASR with CTC on the characters to obtain alignment.\footnote{Based on the released source code, the WhisperX model, claimed to be fine-tuned with phones in \cite{bain2023whisperx}, is in fact fine-tuned with characters. The fact is also noted in \cite{rousso2024tradition}.}
Alignments are obtained by aligning Whisper predictions with the CTC model \cite{kurzinger2020ctc}.
We reproduce WhisperX on our selected datasets using their released code\footnote{https://github.com/m-bain/whisperX}. 
Finally, we perform a modified version of GradScore~\cite{schmitt2025conformer} on Whisper, computing the gradients of the target token logit with respect to the encoder output for cross attention. After obtaining the gradient norm, the same DTW algorithm is used to extract the alignment.

\subsection{Evaluation metric}

We use the strict evaluation protocol \cite{strgar2023phoneme} and measure $\text{F}_1$ comparing the predicted and the ground truth boundaries.
We deem a hypothesized boundary as a true positive when the ending timestamp of a word falls within a given tolerance of a ground truth word and the word identities match, stricter than the evaluation used in \cite{rousso2024tradition}.

\begin{table*}
\centering
\footnotesize
\caption{$\text{F}_1$ scores with 50 ms and 100 ms tolerance comparing different sets of attention heads and output tokens. Oracle heads are the single best heads identified with ground truth boundaries. Whisper used to average the upper half of the decoder layers to produce alignments. The new default (after commit \texttt{dd985ac}) is based on a fixed set of heads. All settings are evaluated on Whisper \textit{medium}.}
\begin{tabular}{llllcccccc}
\toprule
&  & & \multicolumn{2}{c}{\textbf{\textbf{TIMIT}}} &\multicolumn{2}{c}{\textbf{LS}} &\multicolumn{2}{c}{\textbf{\textbf{AMI}}} \\
\cmidrule(lr){4-5}\cmidrule(lr){6-7}\cmidrule(lr){8-9}
\textbf{Model} & \textbf{Heads} & \textbf{Token type} &\textbf{50ms} &\textbf{100ms} &\textbf{50ms} &\textbf{100ms} &\textbf{50ms} &\textbf{100ms}\\
\midrule
Whisper & oracle & wordpieces & 76.0 & 93.4 & 69.6 & 89.6 & 67.1 & 79.8 \\
Whisper & oracle & character & 90.4 & 96.3 & 85.4 & 95.4 & 75.2 & 83.1 \\
\midrule
Whisper & averaging upper half layers & wordpieces & 64.0 & 87.6 & 62.5 & 87.3 & 47.1 & 67.8 \\
Whisper  & fixed heads (commit \texttt{dd985ac}) & wordpieces & 41.2 & 67.1 & 39.8 & 66.6 & 28.5 & 54.6 \\
Whisper & proposed norm filtering (top 10) & character & 80.7 & 94.7 & 80.6 & 93.4 & 61.9 & 77.4 \\
\bottomrule
\end{tabular}
\label{tab:baselines}
\end{table*}

\begin{table*}
\parbox{.65\linewidth}{
\centering
\footnotesize
\caption{$\text{F}_1$ scores with 50 ms and 100 ms tolerance comparing the proposed approach with other aligners.}
\begin{tabular}{llllcccccc}
\toprule
&  & & \multicolumn{2}{c}{\textbf{\textbf{TIMIT}}} &\multicolumn{2}{c}{\textbf{LS}} &\multicolumn{2}{c}{\textbf{\textbf{AMI}}} \\
\cmidrule(lr){4-5}\cmidrule(lr){6-7}\cmidrule(lr){8-9}
\textbf{Model} & \textbf{Aligner} & \textbf{Token type} &\textbf{50ms} &\textbf{100ms} &\textbf{50ms} &\textbf{100ms} &\textbf{50ms} &\textbf{100ms}\\
\midrule
MFA & HMM & phone & 91.0 & 98.0 & - & - & - & - \\
WhisperX  & CTC  & character & 79.9	& 91.2 & 79.5 & 89.1 & 63.5 & 74.2 \\
Whisper  & gradient norm & wordpiece & 63.2& 86.6 & 53.4 & 79.4 & 44.5 & 64.6 \\
\midrule
Whisper & oracle heads & character & 90.4 & 96.3 & 85.4 & 95.4 & 75.2 & 79.8 \\
Whisper & norm-filtered heads & character & 80.7 & 94.7 & 80.6 & 93.4 & 61.9 & 77.4 \\
\bottomrule
\end{tabular}
\label{tab:baselines2}
}
\hfill
\parbox{.275\linewidth}{
\footnotesize
\centering
\caption{$\text{F}_1$ scores with 50 ms comparing CrisperWhisper (Crisper) and the proposed approach.}
\begin{tabular}{lcc}
\toprule
 & \textbf{Crisper} & \textbf{Proposed} \\
\midrule
TIMIT   & 74.0 & 83.3 \\
LS      &  76.7 & 80.4 \\
AMI     & 64.9 & 65.3 \\
\bottomrule
\end{tabular}
\label{tab:crisper_vs_whisperalign}
}
\end{table*}

\section{Results and Discussions}

We first study the first hypothesis, trying to find attention maps that represent alignments.
We then study how using characters with teacher forcing makes the attention maps finer.
We compare comprehensively to other models, and test whether our findings generalize to other seq2seq models.

\subsection{Internal word aligners in Whisper}
\label{sec:results:topline}

We first look for attention maps that represent word alignments, and how they distribute over the decoder layers.
We select the head that gives the highest $\text{F}_1$ score using the ground truth, and we refer to this head as the \textit{oracle head}.
Note again that we might find different oracle heads for different input utterances. 

As shown in Table \ref{tab:baselines}, the oracle heads show strong performance across all tolerance level.
In fact, as we will see in Table \ref{tab:baselines2}, the performance from oracle heads is close to the ones from MFA.
This indicates the strong potential of using Whisper as a word aligner.
We further examine the distributions of oracle heads across different decoder layers in Whisper \textit{medium}.
Figure \ref{fig:freq-best-head} shows that most of the best alignments are obtained from only a few specific heads. The top 20 frequent oracle heads appear in 72\% of the samples in the training set of TIMIT when using wordpieces (and 95\% when using characters).
In other words, most of the heads (384 in total) are in fact not useful for alignments.
In addition, we can see from Figure \ref{fig:freq-best-head} that not all oracle heads are located in layers close to the output.

\begin{figure}[h!]
\includegraphics[width=\linewidth]{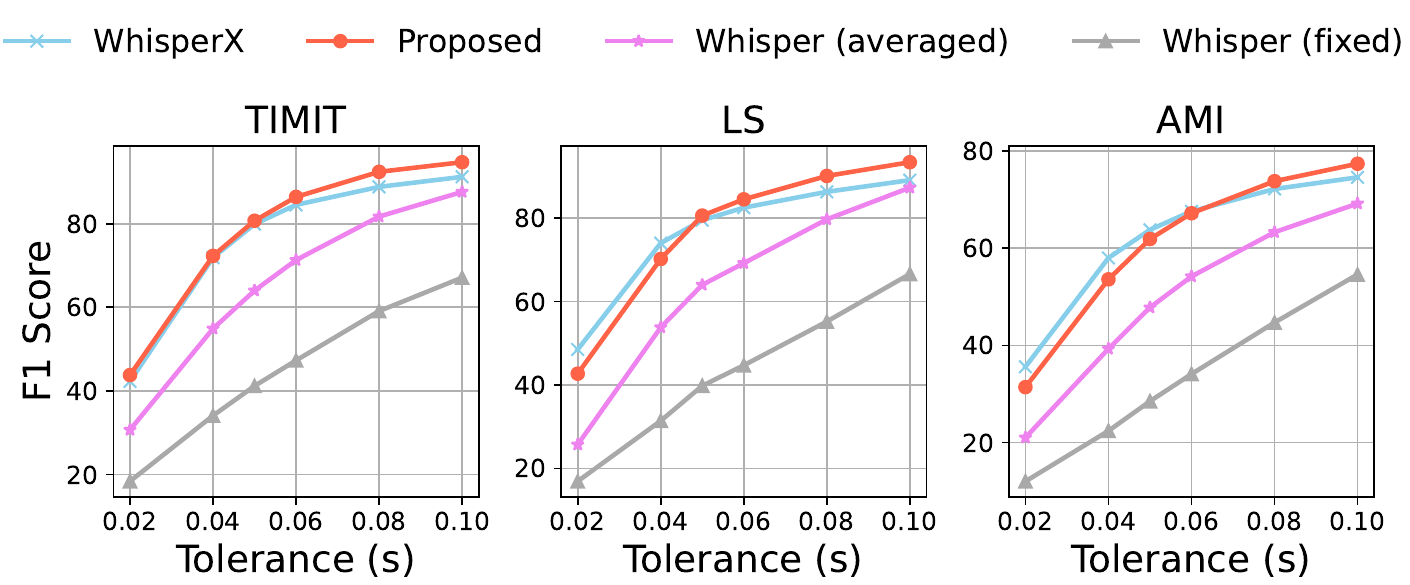}
\caption{$\text{F}_1$ 
scores at different tolerance levels. Except for WhisperX, which uses CTC to derive word boundaries, all other approaches are based on cross-attention maps in Whisper \textit{medium}. Whisper (averaged) is the approach listed in the third row of Table~\ref{tab:baselines}}
\label{fig:tolerance}
\end{figure}

\subsection{Comparison to other aligners}

\label{sec:results:aligner-compare}
We present our main results comparing other aligners with the proposed attention filtering (keeping the top 10 heads) and character-level teacher forcing. 
The derived alignments are then converted into word segments by grouping characters within a complete word. 
We reproduce previous work under the same evaluation pipeline, using Whisper \textit{medium} to ensure consistency.
We also present alignment performance under various levels of tolerances in Figure \ref{fig:tolerance}.

From Table \ref{tab:baselines}, using the proposed approach improves over the original Whisper (commit \texttt{dd985ac}) by a large margin.
With a 50 ms tolerance, our approach achieves 39.5\%, 40.8\% and 33.4\% absolute improvements on TIMIT, LibriSpeech, and AMI, respectively.
The proposed filtering with column and row norms when using characters is sometimes even better than the oracle heads with wordpieces.
We also find that simply averaging heads is better than the pre-defined heads in Whisper.

In Table \ref{tab:baselines2}, we observe that the word timestamps derived from the proposed approach outperforms WhisperX on TIMIT and LibriSpeech, and achieve similar performance on AMI. 
This eliminates the need for an extra aligner (wav2vec 2.0 CTC) to extract word timestamps \cite{bain2023whisperx},
showing the effectiveness of filtering attention heads.
Our approach is also on par with MFA within a tolerance of 100 ms, but only 10.3\% behind with a tolerance of 50 ms.
The gap can be further reduced if we are able to filter heads carefully. The challenges of filtering heads will be discussed in \S\ref{sec:challenges}.
While using the gradient norms of Whisper to align eliminates the need for head filtering, it is generally behind most approaches and requires further exploration. 

Lastly, we compare our approach to CrisperWhisper \cite{wagner2024crisperwhisper}\footnote{https://github.com/nyrahealth/CrisperWhisper}, where they fine-tune Whisper on several spontaneous speech datasets, including AMI. 
CrisperWhisper uses a supervised attention loss similar to~\cite{9746310} to enforce the correspondence between attention heads and ground truth alignments. 
Again, to ensure the performance gap is only attributed to the quality of alignments, we align on the transcriptions predicted by CrisperWhisper for evaluating our approach.
As reported in Table \ref{tab:crisper_vs_whisperalign},
our approach is consistently better than CrisperWhisper, with up to 9.3\% improvement on TIMIT, while achieving comparable performance on AMI, despite CrisperWhisper being fine-tuned on it.

\subsection{Factors that affect word-level alignment accuracy}
\label{sec:results:factors}
Given the strong performance of the proposed approach, we analyze different factors that affect attention-based alignments with qualitative and quantitative evidence.

\begin{figure}
    \centering
    \includegraphics[width=0.9\linewidth]{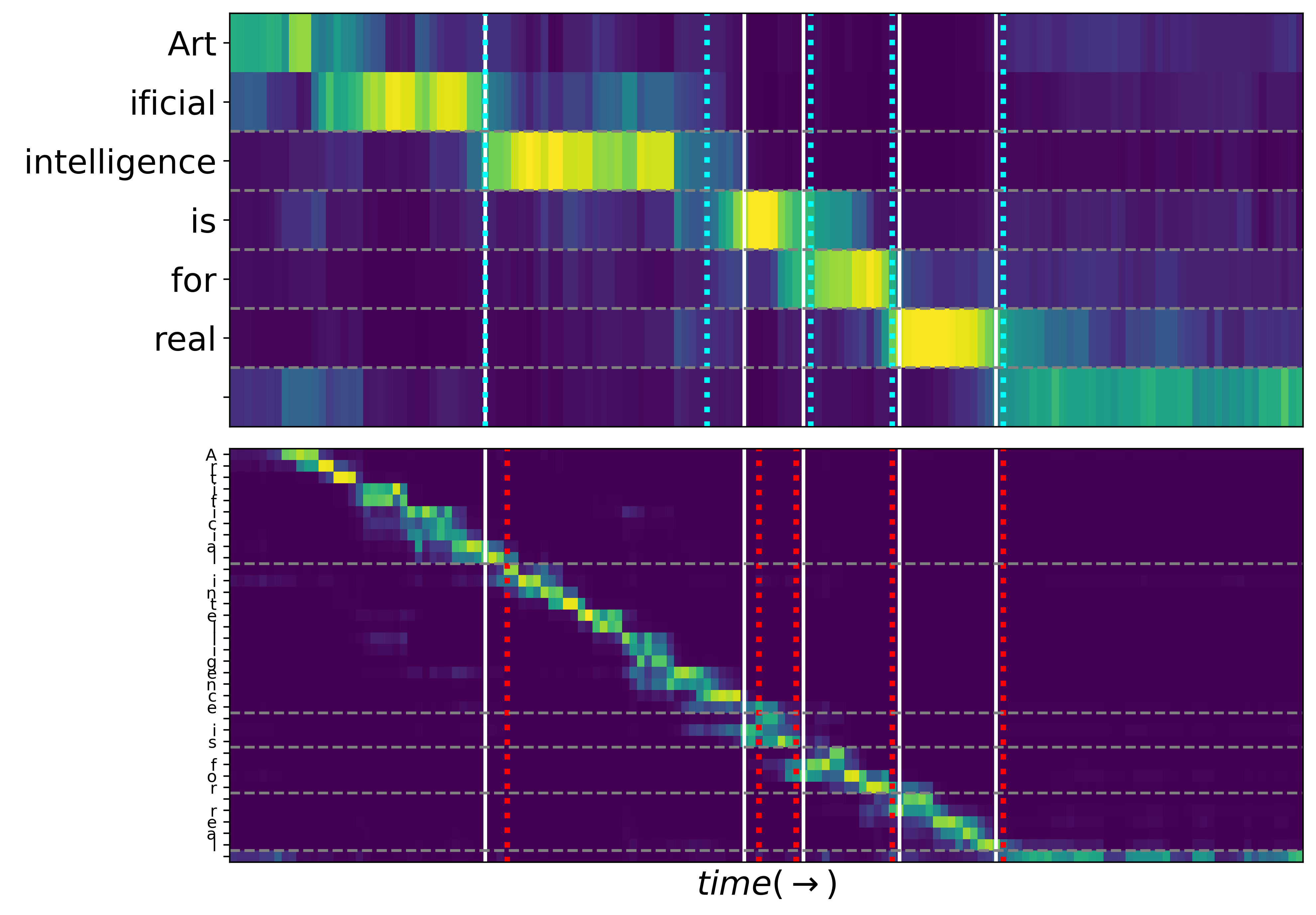}
    \caption{The oracle head attention map using wordpieces (top) and characters (bottom) on an example utterance in TIMIT. The white solid lines are the ground truth, and the dotted lines are the predicted timestamps.}
    \label{fig:best-head}
\end{figure}

\begin{figure}
    \centering
    \includegraphics[width=0.9\linewidth]{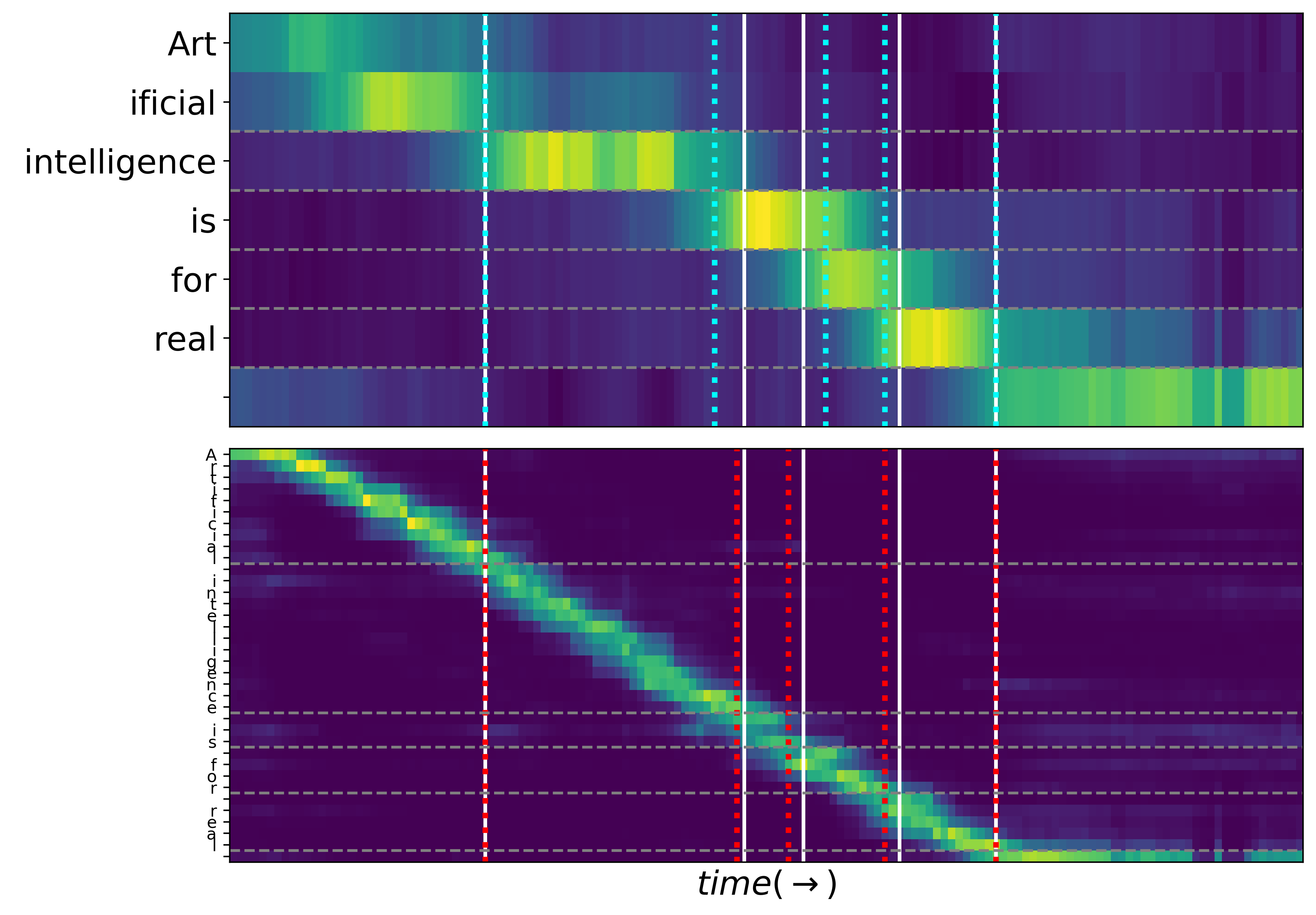}
    \caption{A comparison of averaging attention heads in the upper half of the decoder layers when using wordpieces (top; Whisper's old approach) to our norm-filtered heads when using characters (bottom). The white solid lines are the ground truth, and the dotted lines are the predicted timestamps.}
    \label{fig:our-approach}
\end{figure}

\begin{figure}
    \centering
    \includegraphics[width=\linewidth]{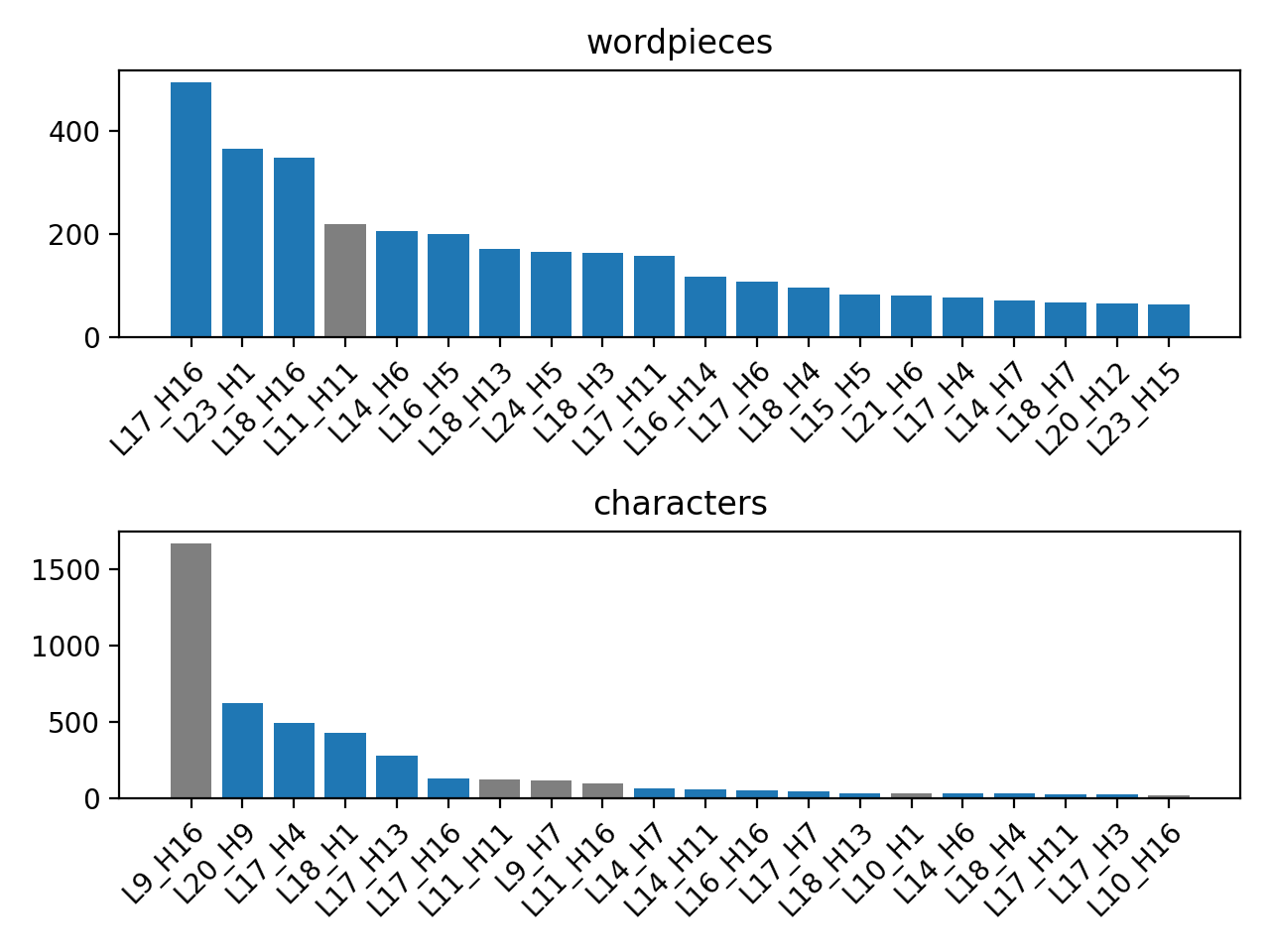}
    \caption{Top 20 frequent oracle heads of Whisper \textit{medium} on TIMIT with wordpieces (top) and characters (bottom).
    The y-axis is the number of testing samples. The x-axis shows the top 20 frequent oracle heads, labeled by its layer index and head index
    There are 24 decoder layers, with 16 heads in each layer. Blue indicates the head is in the upper half of the decoder layers, and grey indicates that the head is within the lower half.}
    \label{fig:freq-best-head}
\end{figure}

\subsubsection{Output token types}
\label{sec:token-granularity}
To show that alignments using characters can be more fine-grained than those using wordpieces,
we first visualize the normalized attention maps aligned with wordpieces or characters. 
As shown in Figure~\ref{fig:best-head}, the alignments from the oracle heads using wordpieces are more contextualized than using characters.
The difference is more obvious in 
Figure~\ref{fig:our-approach}, where the cost matrix is computed after averaging selected attention maps. Clearly, using characters is sharper and more monotonic.

In addition to the qualitative analysis in Figure \ref{fig:best-head} and \ref{fig:our-approach}, we present the importance of token size.
Approaches aligning with smaller tokens, such as phones and characters are in general better than aligning with larger wordpieces, as shown in Table \ref{tab:baselines}.
The results indicate that using smaller tokens, such as characters, can effectively limit the number of frames a token is associated with in cross attention, encouraging more accurate correspondence of the input and output.

\begin{figure}
\centering
\includegraphics[width=4.3cm]{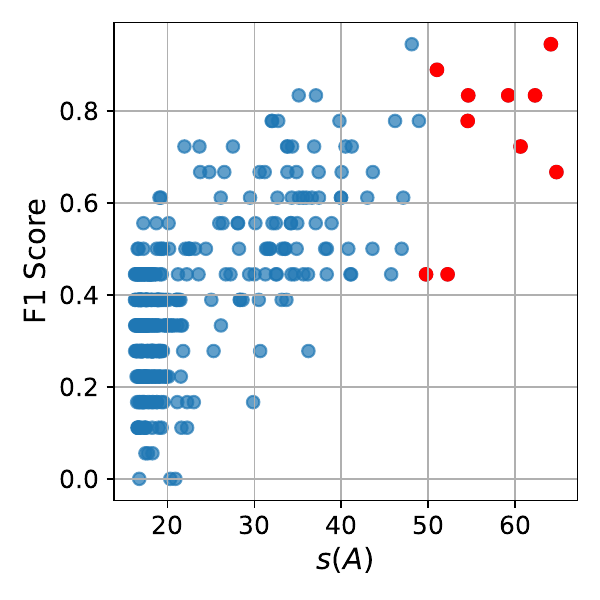}
\includegraphics[width=4.3cm]{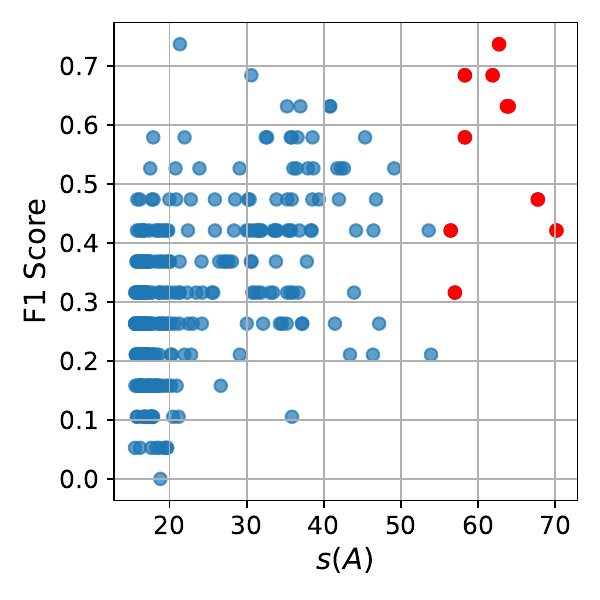}
\caption{$\text{F}_1$ scores of 384 heads against norm-based score in Eq \eqref{eq:norm_score} for two sample utterances. The selected top 10 heads are marked in red. In these two example utterances and consistently in others, heads with higher scores tend to have higher $\text{F}_1$ scores.}
\label{fig:scores-f1}
\end{figure}

\subsubsection{Characteristics of oracle heads}
\label{sec:challenges}

While in \S\ref{sec:results:topline}, we have shown that oracle heads can be identified using the ground truth,
but finding such heads without the ground truth is challenging.
Especially, there are multiple heads that seem to be representing alignments but do not all align with word boundaries.

To see whether the proposed approach effectively captures the oracle head using characters, we show the $\text{F}_1$ scores of individual heads against the proposed attention score in Figure~\ref{fig:scores-f1}. 
We observe that heads with higher scores in general tend to include the oracle head and have higher $\text{F}_1$ scores.
We further show in Table~\ref{tab:char_topk_results} 
oracle heads are frequently included in the top 10 heads based on our filtering criterion.
Keeping more heads after a certain point degrades performance as it starts to include heads that do not represent alignments.

\subsubsection{Filtering heads dynamically}

Contrary to the common practice of using a fixed set of heads for identifying alignments (e.g., the default Whisper and CrisperWhisper), we do not find a small set of heads that consistently represent alignments, especially when using characters as output tokens. 
Using a fixed set of heads also does not seem to generalize well across datasets as shown in  Table \ref{tab:baselines} and Table \ref{tab:crisper_vs_whisperalign}.
Choosing a set of heads independently for every individual utterance leads to the best performance.

In Figure~\ref{fig:freq-best-head}, we show where the heads that represent alignments tend to occur. The top 20 oracle heads tend to appear in the upper half of the decoder layers when using wordpieces, but there does not seem to be a clear pattern when using characters.
This again shows the importance of filtering heads dynamically based on individual utterances.

\begin{table}
\centering
\caption{A comparison of $\text{F}_1$ scores when keeping different amounts of attention heads and how often the oracle head is included (hit rate).}
\begin{tabular}{lcccc}
\toprule
 & \multicolumn{2}{c}{\textbf{TIMIT}} & \multicolumn{2}{c}{\textbf{AMI}} \\
\cmidrule(lr){2-3}\cmidrule(lr){4-5}
  & \textbf{$\text{F}_1$ (50 ms)} & \textbf{Hit rate (\%)} & \textbf{$\text{F}_1$ (50 ms)} & \textbf{Hit rate (\%)} \\
\midrule
Oracle  & 90.7  & -    & 75.2  & -     \\
Top 1   & 35.2  & 0.0  & 43.5  & 0.0   \\
Top 5   & 78.2  & 49.5 & 62.8  & 61.3  \\
Top 10  & 80.7  & 77.1 & 61.9  & 82.5  \\
Top 20  & 75.0  & 83.2 & 58.6  & 88.9  \\
All     & 42.4  & 100.0  & 28.0  & 100.0   \\
\bottomrule
\end{tabular}
\label{tab:char_topk_results}
\end{table}

\begin{table}
\centering
\footnotesize
\caption{A comparison of different filtering criteria when using characters on TIMIT.}
\begin{tabular}{lc}
\toprule 
\textbf{Filtering criterion} & \textbf{$\text{F}_1$ (50ms)} \\
\midrule
column norm + row norm  & 80.7 \\
column norm  & 80.2\\
row norm & 80.5 \\
row entropy  & 80.1 \\ 
coverage & 42.2 \\
\bottomrule
\end{tabular}
\label{tab:selection-approaches}
\end{table}

\subsubsection{Head filtering criteria}
As discussed in \S\ref{sec:method:head-select}, there are several criteria possible for filtering heads.
In Table~\ref{tab:selection-approaches}, we study how the criteria impact the performance, comparing norms, entropy, and coverage.
Using the sum of column and row norms gives the highest $\text{F}_1$ score, while using just the column norm or the row norm are both strong as well.  
Using the row entropy also works well as expected, given the connection between the $\ell_2$ norm and the R\'enyi entropy discussed in \S\ref{sec:method:head-select}.
Measuring the coverage penalty~\cite{klein2018opennmt} is empirically less effective and depends on a threshold.

\subsection{Transferability of the approach to other seq2seq models}

The proposed approach is not limited to Whisper medium, and it can be applied to other seq2seq models.
In this section, we apply our approach to a deeper Whisper model, large-v2 and another large-scale model, Canary-1B~\cite{puvvada2024less}. Canary-1B is a multilingual ASR and speech translation model, having 24 layers in both the encoder and decoder, matching the size of Whisper \textit{medium}. Canary-1B differs from Whisper in several ways: the vocabulary size of the Canary model is comparatively small (size of 1024 for each language), so that the wordpieces are smaller in size and do not cover complete words as often. The Canary model performs aggressive subsampling in the encoder, having a frame rate of 80 ms in the encoder features, whereas Whisper has a small frame rate of 20 ms.
The frame rate may significantly affect the resolution of the alignment, so Canary-1B is not expected to perform well, especially when evaluating under a stricter tolerance.

In Table~\ref{tab:model_comparison}, we report the result of averaging the upper half of the decoder layers for wordpieces as it gives stronger performance. For characters, we sweep across \{5, 10, 15, 20\} heads to select and report the best result. 
From Table~\ref{tab:model_comparison}, we see that our findings transfer well to Whisper large-v2, though the absolute performance is slightly worse than Whisper medium.
The findings also transfer to Canary-1B, though the absolute performance is generally lower due to the inherent large frame rate of 80ms.
When evaluating with a tolerance of 100 ms, the performance with characters (in the oracle case, for example) goes from 63.8\% to 84.1\%, confirming the fact that there also exist attention maps that represent alignments in Canary-1B and that frame rate is the limiting factor in this case for extracting alignments.

\begin{table}
\centering
\footnotesize
\caption{Generalization of our findings to other seq2seq models}
\begin{tabular}{lcc}
\toprule
                    & wordpieces   & characters \\
\midrule
Whisper medium      & 64.0         & 80.7      \\
Whisper large-v2    & 58.4         & 78.9      \\
\midrule
Canary-1B           & 32.7         & 35.4      \\
Canary-1B (oracle)  & 64.8         & 63.8      \\ 
\bottomrule
\end{tabular}
\label{tab:model_comparison}
\end{table}

\section{Conclusion}

In this work, we show the existence of attention heads that represent alignments in the Whisper decoder.
We propose a filtering approach to automatically discover those heads.
We also show that using characters instead of wordpieces produces finer attention maps, resulting in finer alignments.
The proposed approach outperforms recent work on Whisper-based alignments by a large margin, and is better than WhisperX in most settings.
These internal word aligners also exist in other large-scale seq2seq models, which we hope to further study in the future.

\bibliographystyle{IEEEbib}
\bibliography{refs}
\end{document}